\documentclass[%
    aps,
    prd,
    reprint,
    superscriptaddress,
    floatfix,
    nofootinbib,
    noamsmath,
    noamssymb
]{revtex4-2}

\usepackage[T1]{fontenc}
\usepackage[utf8]{inputenc}
\usepackage{graphicx}
\usepackage{xcolor}
\usepackage{mathtools}
\usepackage{amssymb}
\usepackage{lmodern}
\usepackage{bm}
\usepackage{siunitx}
\usepackage[final]{microtype}
\usepackage{hyperref}

\newcommand{\mytitle}{From imaginary to real chemical potential QCD with functional methods}

\newcommand{\JLU}{%
    Institut f\"{u}r Theoretische Physik,
    Justus-Liebig-Universit\"{a}t Gie\ss{}en,
    35392 Gie\ss{}en,
    Germany
}

\newcommand{\HFHF}{%
    Helmholtz Forschungsakademie Hessen f\"{u}r FAIR (HFHF),
    GSI Helmholtzzentrum f\"{u}r Schwerionenforschung,
    Campus Gie\ss{}en,
    35392 Gie\ss{}en,
    Germany
}

\hypersetup{%
    pdftitle = {\mytitle},
    pdfauthor = {Julian Bernhardt, Christian S. Fischer},
    bookmarksopen = true,
    colorlinks = true,
    linkcolor = green!40!black,
    citecolor = blue!50!black,
    urlcolor = blue!50!black
}

\DeclareSIUnit{\MeV}{\mega\electronvolt}
\DeclareSIUnit{\GeV}{\giga\electronvolt}
\DeclareSIUnit{\fm}{\femto\meter}

\DeclareMathOperator{\Tr}{Tr}

\renewcommand*{\vec}[1]{\bm{#1}}
\newcommand*{\+}{\hspace*{.08335em}}

\newcommand*{\dd}{\mathrm{d}}
\newcommand*{\ii}{\mathrm{i}}

\newcommand*{\bbZ}{\mathbb{Z}}

\newcommand*{\Tc}{T_{\textup{c}}}
\newcommand*{\Nc}{N_{\textup{c}}}

\newcommand*{\gs}{g}
\newcommand*{\massu}{m_{\textup{u}}}
\newcommand*{\massd}{m_{\textup{d}}}
\newcommand*{\masss}{m_{\textup{s}}}
\newcommand*{\muu}{\mu_{\textup{u}}}
\newcommand*{\mud}{\mu_{\textup{d}}}
\newcommand*{\mus}{\mu_{\textup{s}}}
\newcommand*{\muB}{\mu_{\textup{B}}}

\newcommand*{\cond}{\langle \bar{\psi} \psi \rangle}

\definecolor{dgreen}{rgb}{0.1,0.5,0.1}
\definecolor{lblue}{rgb}{0.2,0.35,1}

\begin{document}

\title{\mytitle}

\author{Julian Bernhardt}
\email{julian.bernhardt@physik.uni-giessen.de}
\affiliation{\JLU}
\affiliation{\HFHF}

\author{Christian S.~Fischer}
\email{christian.fischer@theo.physik.uni-giessen.de}
\affiliation{\JLU}
\affiliation{\HFHF}

\begin{abstract}
We investigate the quality of the extrapolation procedure employed in
Ref.~\cite{Borsanyi:2020fev} to extract the crossover line at real chemical
potential from lattice data at imaginary potential. To this end we employ a
functional approach that does not suffer from the sign problem. We utilize a
well-studied combination of lattice Yang--Mills theory with a truncated set of
Dyson--Schwinger equations in Landau gauge for $2 + 1$ quark flavors. This
system predicts a critical endpoint at moderate temperatures and rather large
(real) chemical potential with a curvature of the pseudo-critical transition line 
comparable to recent lattice extrapolations. We determine the light quark condensate 
and chiral susceptibility at imaginary chemical potentials and perform an analytic
continuation along the lines described in \cite{Borsanyi:2020fev}. We find that
the analytically continued crossover line agrees very well (within one percent)
with the explicitly calculated one for chemical potentials up to about 80 \% of
the one of the critical end point. The method breaks down in the region where
the chiral susceptibility as a function of the condensate cannot any longer be
well described by a polynomial. 
\end{abstract}

\maketitle

\section{\label{introduction}%
    Introduction
}

One of the major goals in studying the QCD phase diagram is to extract a
quantitative understanding of the existence, the location and the properties of
a critical endpoint (CEP). Observables connected to the appearance of the CEP
have been probed in heavy-ion-collision experiments at RHIC/BNL
\cite{Bzdak:2019pkr} and HADES (FAIR Phase-0) \cite{Salabura:2020tou}, and will
be explored in detail in the future CBM/FAIR experiment \cite{Friman:2011zz}.

From a theoretical perspective, a high precision study of the location of the
CEP is hampered by two very different problems. In lattice QCD, the fermion sign
problem prohibits direct calculations at real chemical potential, see e.g.
\cite{Nagata2022} for a review. Extrapolation procedures from zero chemical potential 
include re-weighting \cite{Hasenfratz:1991ax,Fodor:2001au,Fodor:2001pe,Giordano:2020roi} 
and Taylor expansion schemes 
\cite{Allton:2002zi,Gavai:2008zr,Borsanyi:2012cr,Bazavov:2017dus,Giordano:2019gev,Bazavov2019,Bazavov:2020bjn,Bollweg2022}
allowing for indirect access to important quantities such as pseudo-critical 
transition temperatures, equations of state and fluctuations of conserved charges 
at moderate chemical potential \cite{Ratti:2022qgf}. In addition, direct lattice 
simulations at imaginary chemical potential provide a basis for a number of extrapolation
procedures into the real chemical potential domain based on the analytic properties
of the QCD partition function. This procedure has been pioneered by de Forcrand and Philipsen 
\cite{deForcrand:2002hgr} and has been developed since, see e.g.
\cite{Borsanyi:2020fev,Borsanyi:2021sxv,Dimopoulos:2021vrk,Borsanyi:2022soo,Borsanyi2022a}
 and Refs. therein. 
 
 On the other hand, functional approaches to the CEP, i.e.,
approaches via Dyson--Schwinger equations (DSE) and/or the functional
renormalization group (FRG), do not suffer from the sign problem and allow for
a mapping of the whole QCD phase diagram, see e.g. \cite{Fischer:2018sdj} for a
review and
\cite{Fischer:2014ata,Isserstedt:2019pgx,Fu:2019hdw,Gao:2020qsj,Gao:2020fbl,Gunkel:2021oya}
 for recent results. However, they inherently depend on approximations and
truncations which have to be carefully controlled either in combined approaches
or by comparison with lattice results in controlled environments where both
approaches are applicable, such as imaginary chemical potential. This work is
devoted to the latter. We revisit a truncation scheme of Dyson--Schwinger
equations that has been used in Ref.~\cite{Gunkel:2021oya} to perform
calculations at real chemical potential up to the critical endpoint and apply
it at imaginary baryon chemical potential $\muB$. We perform calculations at
appropriate values of imaginary $\muB$, where direct comparison with
corresponding lattice results from Ref.~\cite{Borsanyi:2020fev} is possible. We
then employ a similar extrapolation procedure as on the lattice and compare the
extrapolated results for the crossover transition line with the ones explicitly
calculated in our functional approach. Our results may serve as a quality check
for the extrapolation method used on the lattice.

The paper is organized as follows. In the next section, we briefly revisit our
framework. We summarize the truncation scheme of Ref.~\cite{Gunkel:2021oya} in
Section \ref{DSE-framework} and review the extrapolation procedure from
imaginary to real chemical potential employed in Ref.~\cite{Borsanyi:2020fev} in
Section \ref{imu-framework}. In Section \ref{results}, we present and discuss
our results. We summarize and conclude in Section \ref{summary}.

\section{\label{framework}%
    Framework
}

\subsection{\label{DSE-framework}%
	Truncated set of coupled Dyson--Schwinger equations
}

The central quantity for all following investigations is the dressed quark
propagator $S_{f}$ of quark flavor $f$ at non-zero temperature $T$ and quark
chemical potential $\mu_{f}$. Its inverse is parametrized as\footnote{%
    We work in Euclidean space-time with positive metric signature (++++). The
    Hermitian $\gamma$-matrices satisfy $\{\gamma_{\nu}, \gamma_{\rho}\} = 2
    \delta_{\nu \rho}$.%
}%
\begin{equation}
    \label{eq:quark_propagator}
    S_{f}^{-1}(p)
    =
    \ii \gamma_{4} \tilde{\omega}_{n}^{f} C_{f}(p)
    +
    \ii \vec{\gamma} \cdot \vec{p} A_{f}(p)
    +
    B_{f}(p)
\end{equation}
where $p = (\vec{p}, \tilde{\omega}_{n})$ labels the four-momentum, while
$\tilde{\omega}_{n}^{f} = \omega_{n} + \ii \mu_{f}$ represents a combination of
the fermionic Matsubara frequencies $\omega_{n} = (2n + 1) \pi T$, $n \in \bbZ$,
with the chemical potential. At this point, we already remark that the
definition of $\tilde{\omega}_{n}^{f}$ universally holds for both real and
imaginary chemical potentials in our framework. Consequently, imaginary chemical
potentials effectively correspond to a real shift of the Matsubara frequencies
\cite{Braun:2009gm,Fischer:2014vxa}. The quark dressing functions $A_{f}$,
$B_{f}$ and $C_{f}$ in Eq.~\eqref{eq:quark_propagator} encode the
non-perturbative momentum dependence of the propagator.

\begin{figure}[t]
    \centering%
    \includegraphics{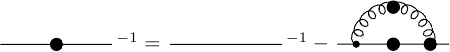}%
    \\[1em]%
    \includegraphics{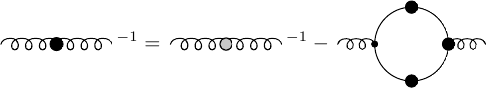}%
    \caption{\label{fig:dses}%
        DSE for the quark propagator (top) and truncated gluon DSE (bottom).
        Large filled dots indicate dressed quantities; solid and curly lines
        represent quark and gluon propagators, respectively. There is a separate
        DSE for each flavor. The large gray dot denotes the quenched gluon
        propagator that is taken from the lattice while the quark loop is
        evaluated explicitly. The latter contains an implicit flavor sum.
    }%
\end{figure}

We obtain the quark propagator as a solution of its corresponding
Dyson--Schwinger equation (DSE) illustrated in Fig.~\ref{fig:dses} and given by
\begin{equation}
    \label{eq:quark_dse}
    S_{f}^{-1}(p)
    =
    Z_{2}
    \bigl(
        \ii \gamma_{4} \tilde{\omega}_{n}^{f}
        +
        \ii \vec{\gamma} \cdot \vec{p}
        +
        Z_{m} m_{f}
    \bigr)
    -
    \Sigma_{f}(p)
    \,,
\end{equation}
with $Z_{2}$ and $Z_{m}$ labelling the wave function and mass renormalization
constants, respectively, which are calculated in vacuum using a
momentum-subtraction scheme. Additionally, $m_{f}$ denotes the flavor-dependent
current quark mass. The quark self-energy reads
\begin{multline}
    \label{eq:quark_self-energy}
    \Sigma_{f}(p)
    =
    (\ii \gs)^{2}
    \frac{4}{3}
    \frac{Z_{2}}{\tilde{Z}_{3}}
    T
    \sum_{\omega_{n}}
    \int \frac{\dd^{3} q}{(2 \pi)^{3}}
    D_{\nu\rho}(p - q)
    \gamma_{\nu}
    \times
    \\
    \times
    S_{f}(q)
    \Gamma_{\rho}^{f}(p, q; p-q)_{Q}
    \,,
\end{multline}
where $g$ denotes the strong coupling constant, $\tilde{Z}_{3}$ the ghost
renormalization constant and $D_{\nu\rho}$ the dressed gluon propagator. The two
factors of $\ii g$ are pulled out of the vertices such that $\Gamma_{\rho}^{f}$
represents the reduced dressed quark--gluon vertex. The trace over color space,
which was already performed, results in the prefactor of $4/3$ for $\Nc = 3$
colors.

Solving the quark DSE requires knowledge about the gluon $D_{\nu\rho}$ and
the quark-gluon vertex $\Gamma_{\rho}^{f}$. As in Ref.~\cite{Gunkel:2021oya},
we use the truncation for the gluon DSE shown in Fig.~\ref{fig:dses}, where all
Yang--Mills diagrams (one- and two-loop diagrams with ghosts and gluons only)
are replaced by an inverse quenched gluon propagator that is reconstructed
from temperature-dependent fits to results of quenched lattice calculations
\cite{Fischer:2010fx,Maas:2011ez,Eichmann:2015kfa}. Denoting the quenched gluon
by $D_{\nu\rho}^{\textrm{YM}}$, the truncated gluon DSE is given by
\begin{equation}
    \label{eq:gluon_dse}
    D_{\nu\rho}^{-1}(p)
    =
    \bigl[
        D_{\nu\rho}^{\textrm{YM}}(p)
    \bigr]^{-1}
    +
    \Pi_{\nu\rho}(p)
    \,.
\end{equation}
Here, $\Pi_{\nu\rho}$ represents the quark loop that is evaluated explicitly
and reads
\begin{multline}
    \label{eq:quark_loop}
    \Pi_{\nu\rho}(p)
    =
    \frac{(\ii \gs)^{2}}{2}
    \sum_{f}
    \frac{Z_{2}}{\tilde{Z}_{3}}
    T
    \sum_{\omega_{n}}
    \int \frac{\dd^{3} q}{(2 \pi)^{3}}
    \Tr\bigl[
        \gamma_{\nu}
        S_{f}(q)
        \times
        \\
        \times
        \Gamma_{\rho}^{f}(q-p, q; p)_{G}
        S_{f}(q - p)
    \bigr]
    \,,
\end{multline}
where the prefactor of $1/2$ again originates in the color trace. The flavor sum
$f$ runs over $N_f=2+1$ quark flavors, i.e., we work with up, down and strange
quarks. The influence of the charm quark on the QCD phase diagram has been
addressed in \cite{Fischer:2014ata} and found to be negligible. The explicit
inclusion of the quark loop generates an unquenched gluon with contributions
from all active quark flavors. Consequently, the different quarks also influence
each other indirectly via the gluon.

The last quantity we need in order to render the previous set of equations
self-contained is the quark--gluon vertex $\Gamma_{\rho}^{f}$. As in
Ref.~\cite{Gunkel:2021oya}, we employ a slightly different form of the vertex in
the quark self-energy and in the quark loop, indicated by subscripts $Q$ and $G$
in Eqs.~\eqref{eq:quark_self-energy} and \eqref{eq:quark_loop}, respectively.
Our ansatz reads:
\begin{align}
    \Gamma^{f}_{\rho}(p,q;k)_{Q}
    &=
    Z_{2}^{f} \gamma_{\rho} \Gamma(k^{2})
    \,,
    \\[.25em]
    \Gamma^{f}_{\rho}(p,q;k)_{G}
    &=
    \Gamma^{f, \text{BC}}_{\rho}(p,q)
    \Gamma(q^{2} + p^{2})
    \,.
\end{align}
Here,
\begin{multline}
    \Gamma^{f, \text{BC}}_{\rho}(p,q)
    =
    \delta_{\rho i}
    \gamma^{i}
    \frac{A_{f}(p) + A_{f}(q)}{2}
    +
    \\
    +
    \delta_{\rho 4}
    \gamma^{4}
    \frac{C_{f}(p) + C_{f}(q)}{2}
\end{multline}
denotes the leading Dirac tensor structure of the Ball--Chiu vertex
\cite{Ball:1980ay} which incorporates to backcoupling effects of the quarks onto
the vertex, while
\begin{equation}
    \Gamma(x)
    =
    \frac{d_{1}}{d_{2} + x}
    +
    \frac{x}{\Lambda^{2} + x}
    \biggl(
        \frac{\alpha(\nu) \beta_{0}}{4 \pi}
        \ln\bigl(
            x / \Lambda^{2} + 1
        \bigr)
    \biggr)^{2 \delta}
\end{equation}
labels a phenomenological vertex dressing function. It consists of an
enhancement at small momenta (IR) inspired by Slavnov--Taylor identities while
it ensures the correct perturbative behavior of the propagators in the UV at
large momenta. The running coupling at the renormalization point
$\nu=\SI{80}{\GeV}$ reads $\alpha(\nu) = 0.3$, the scales $d_{2} =
\SI{0.5}{\GeV^{2}}$ and $\Lambda = \SI{1.4}{\GeV}$ are fixed to match the ones
in the gluon lattice data, and the anomalous dimension is given by $\delta = -9
\Nc / (44\Nc - 8 N_{f})$, while $\beta_{0} = (11 \Nc - 2 N_{f}) / 3$. We work in
a setup of $N_f=2+1$ quark flavors, i.e., the isospin-symmetric limit of
degenerate up and down quarks ($\massu = \massd$, $\muu = \mud$). Furthermore,
we choose $\mus = 0$ such that the baryon chemical potential is given by $\muB =
3 \muu$. The quark masses have been determined in Ref.~\cite{Gunkel:2019xnh} in
the vacuum using the experimental pion and kaon masses. This leads to values of
$\massu = \SI{1.47}{\MeV}$ and $\masss = \SI{37.8}{\MeV}$ at a renormalization
point of $\SI{80}{\GeV}$. The vertex strength parameter $d_{1}$ for this
truncation has been determined in Refs.~\cite{Gunkel:2019xnh,Gunkel:2021oya} to
obtain a pseudocritical temperature at vanishing chemical potential consistent
with lattice results for the subtracted condensate. In this work, we use a
slightly different value that is adapted to the conventions used in
\cite{Borsanyi:2020fev} for the determination of the pseudocritical temperature
(see below for details). This leads to $d_{1}=\SI{12.71}{\GeV^{2}}$ (instead of
$d_{1}=\SI{12.85}{\GeV^{2}}$ as in \cite{Gunkel:2021oya}) in the quark
self-energy and $\SI{8.49}{\GeV^{2}}$ (same as in \cite{Gunkel:2021oya}) in the
quark loop.

With this input, we solve the set of quark and gluon DSEs self-consistently in
the same numerical setup as described in more detail in
Refs.~\cite{Isserstedt:2019pgx,Isserstedt:2021acw} and thus obtain the quark and
gluon propagators as well as quark number and higher susceptibilities for
arbitrary temperatures and (real or imaginary) chemical potentials.

\subsection{\label{imu-framework}%
	Extrapolating from imaginary to real chemical potential
}

In the following, we detail the extrapolation procedure of Ref.~\cite{Borsanyi:2020fev}
based on results for the up quark condensate $\cond$ and the chiral susceptibility $\chi$
at imaginary chemical potential and adapt it to our DSE framework.

With the solutions for the coupled set of DSEs (displayed in
Fig.~\ref{fig:dses}) at hand, we are able to determine the up-quark condensate
straightforwardly from its propagator via taking appropriate traces. The
corresponding quark susceptibility is defined as its derivative with respect to
the quark mass:
\begin{align}
    \label{eq:condensate}
    \cond(T)
    &=
    3 \+ Z_{2}^{f} Z_{m} T
    \sum_{\omega_{k}}
    \int_{\vec{q}}
    \Tr\bigl[ S_{\textup{u}}(q) \bigr]
    \,,
    \\
    \chi(T)
    &=
    \frac{\partial \cond(T)}{\partial \massu}
    \,.
\end{align}
The quark susceptibility is numerically determined in our approach via a finite-difference 
formula.\footnote{In principle, one could also determine the derivative 
directly, but this is about
	an order of magnitude more involved in terms of CPU-time. We have tested the quality of
	the finite difference method in comparable cases and found it to be accurate on the sub-percent 
	level.} 

The expression for the condensate in Eq.~\ref{eq:condensate} is divergent and
needs to be regularized. Similar to Ref.~\cite{Borsanyi:2020fev}, we define the
regularized condensate and susceptibility in the following way:
\begin{align}\label{cond}
    \cond_{\text{reg}}(T)
    &=
    \bigl[ \cond(T) - \cond(0) \bigr]
    \frac{\massu}{f_{\pi}^{4}}
    \,,
    \\
    \chi_{\text{reg}}(T)
    &=
    \bigl[ \chi(T) - \chi(0) \bigr]
    \frac{\massu^{2}}{f_{\pi}^{4}}
    \,,
\end{align}
where $f_{\pi} = \SI{130.41}{\MeV}$ indicates the pion decay constant in the vacuum.

\begin{figure}[t]
    \centering%
    \includegraphics[scale=1.0]{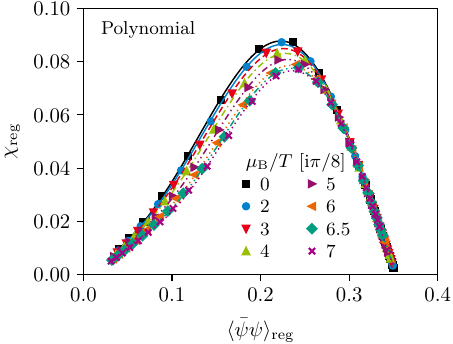}%
    \\[0.5em]
    \includegraphics[scale=1.0]{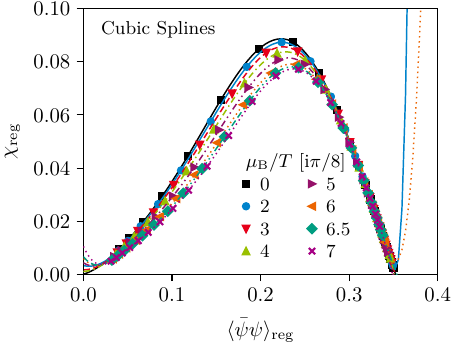}%
    \caption{\label{fig:immu_fits}%
        Chiral susceptibility as a function of the chiral condensate for
        imaginary chemical potentials. The points indicate the DSE data while
        the curves represent polynomial fits (upper panel) or cubic spline
        interpolations (lower panel), respectively.
    }%
\end{figure}

Utilizing these definitions, we determine the pseudocritical transition
temperature as follows:
\begin{enumerate}
    \item We calculate the regularized condensate $\cond(T)$ and susceptibility
    $\chi(T)$ for a discrete set of temperatures at each value of imaginary/real
    chemical potential.
    \item This data is converted to a discrete set of points with the
    dependence $\chi(\cond)$.
    \item We use either a fit to a polynomial of order five or a cubic spline
    interpolation to determine the peak position of these curves. This yields
    the value of the condensate at the pseudocritical temperature $\cond(\Tc)$.
    \item Going back to the discrete set of values for $\cond(T)$ determined in
    step one, we use an appropriate interpolation procedure to extract $\Tc$
    from $\cond(\Tc)$.
\end{enumerate}

\section{\label{results}%
	Results and discussion
}

\begin{figure}[t]
	\centering%
	\includegraphics[scale=0.6]{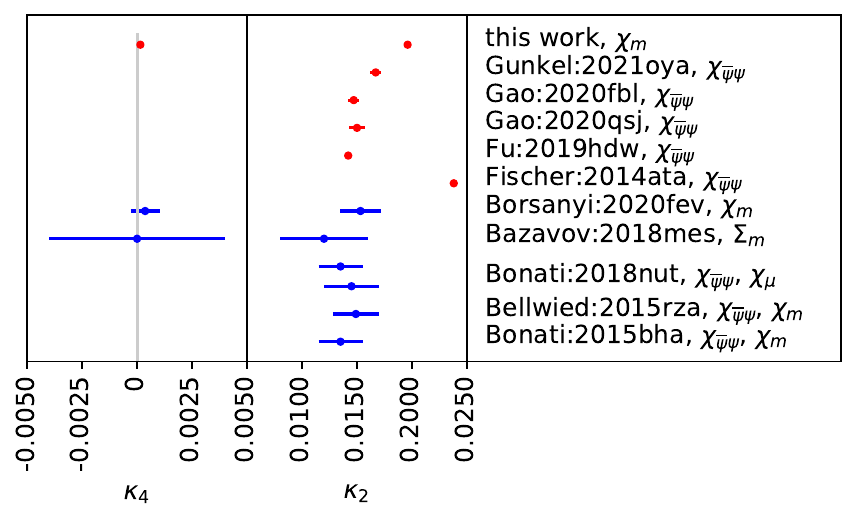}%
	\caption{\label{fig:kappa}%
		Expansion coefficients $\kappa_2$ and $\kappa_4$ for the pseudocritical chiral 
		transition line parametrized in Eq.(\ref{eq:parametrization_Tc}) from different
		sources. Lattice QCD results 
		\cite{Bonati:2015bha,Bellwied:2015rza,Bonati:2018nut,HotQCD:2018pds,Borsanyi:2020fev}
		are in blue, DSE/FRG results \cite{Fischer:2014ata,Fu:2019hdw,Gao:2020qsj,Gao:2020fbl,Gunkel:2021oya}
		in red. 
	}%
\end{figure}

\begin{figure*}[t]
	\centering%
	\includegraphics[scale=1.0]{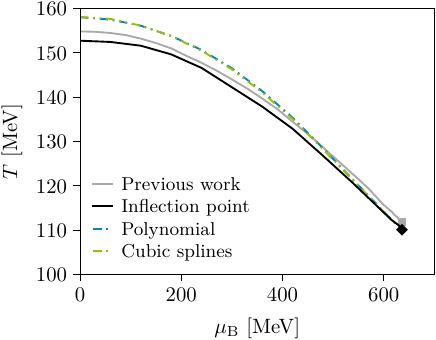}\hfill \includegraphics[scale=0.3]{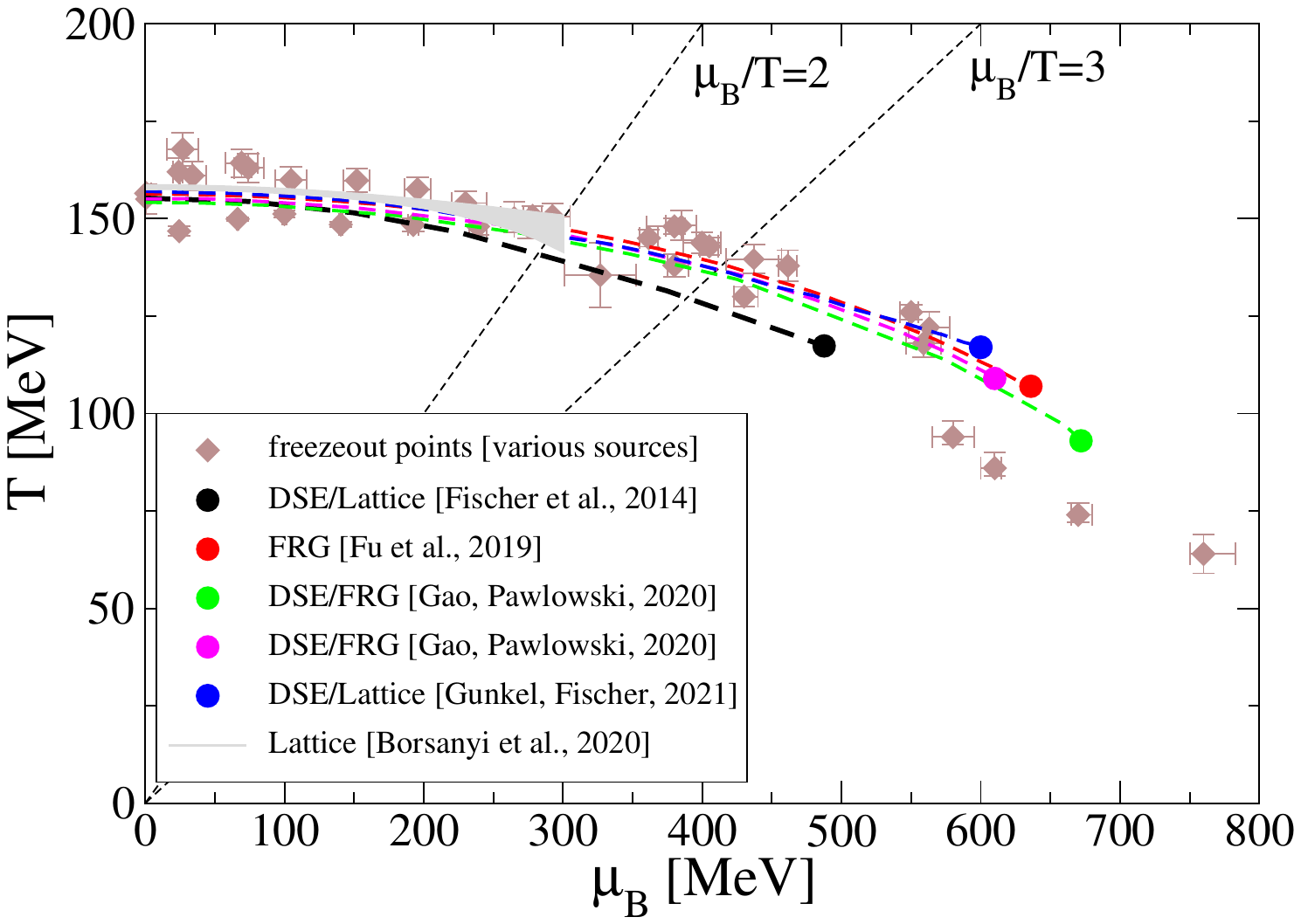}
	\\[0.5em]
	\includegraphics[scale=1.0]{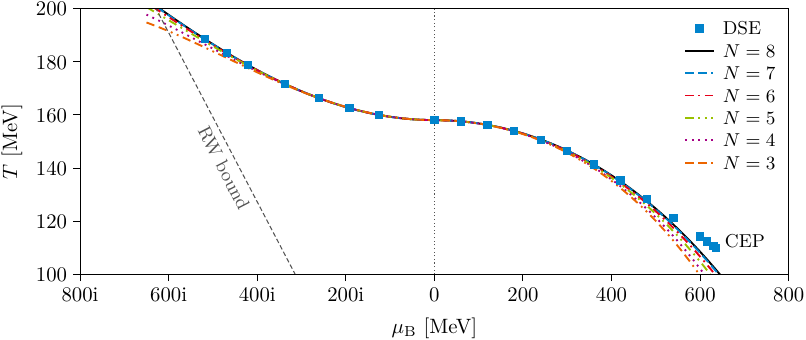}%
	\\[0.5em]
    \includegraphics[scale=1.0]{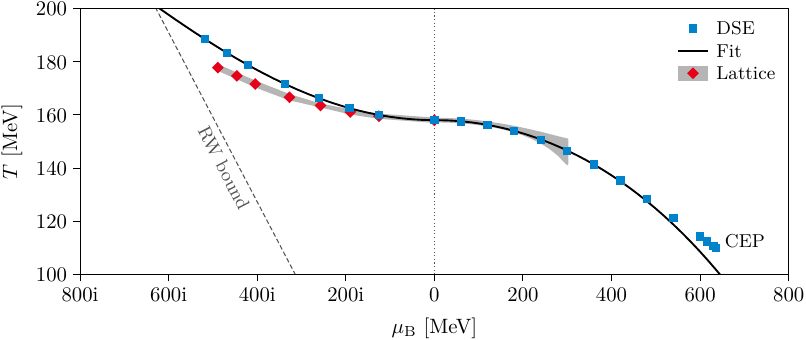}%
	\caption{\label{fig:both_Tc}%
		Upper left diagram: Phase diagram for different definitions of the
		pseudocritical temperature (see main text for explanations).
		Upper left diagram: Phase diagram with critical end point determined in different truncation
		schemes for DSEs and FRGs \cite{Fischer:2014ata,Fu:2019hdw,Gao:2020qsj,Gao:2020fbl,Gunkel:2021oya}.
	    Middle diagram:
	    Results for fits to different numbers of point at imaginary potential. $N$ counts the number
	    of points starting with 1 at zero chemical potential and adding point by point in the direction
	    of large imaginary chemical potential. 
		Lower diagram:
		Pseudocritical transition temperature at both imaginary and real
		chemical potentials from DSEs (points) and fit (line). The gray band
		and the red diamonds correspond to the lattice results from Ref.~\cite{Borsanyi:2020fev}.
	}%
\end{figure*}

\begin{figure*}[t]
	\centering%
	\includegraphics[scale=1.0]{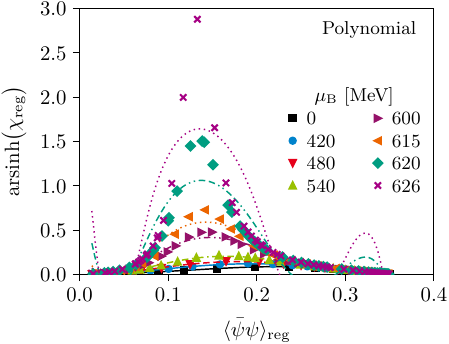}%
	\includegraphics[scale=1.0]{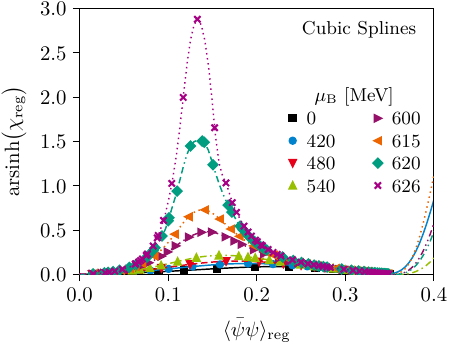}%
	\caption{\label{fig:remu_fits}%
		Chiral susceptibility as a function of the chiral condensate for
		real chemical potentials near the CEP. The points indicate the DSE data
		while the curves represent polynomial fits (left panel) or cubic spline
		interpolations (right panel), respectively.
	}%
	\vspace{1.5em}
	\centering%
	\includegraphics[scale=1.0]{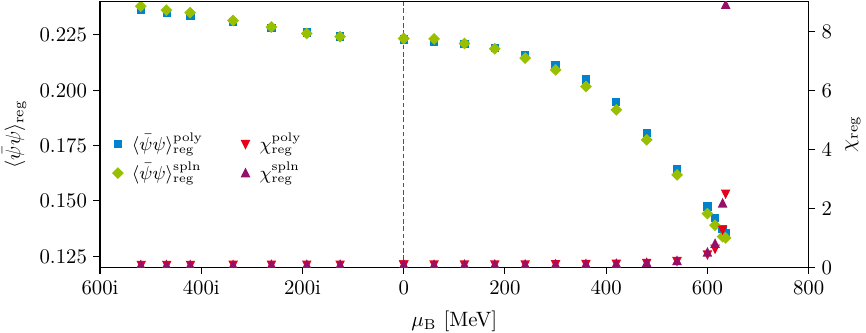}%
	\caption{\label{fig:both_reg}%
		Condensate (blue and green) and susceptibility (red and purple) at the
		pseudocritical temperature for both imaginary and real chemical
		potentials obtained with both polynomial fits and cubic spline
		interpolations.
	}%
\end{figure*}

In Fig.~\ref{fig:immu_fits}, we display the datasets and resulting curves of
$\chi(\cond)$ obtained from steps 2 and 3 (defined at the end of the last
section) for imaginary chemical potentials in the range $\muB / T =( 0, \dots,
7) \times \ii \pi / 8$. The points indicate DSE data, while the curves represent
polynomial fits in the upper panel and cubic spline interpolations in the lower
panel, respectively.

We first notice that both the polynomial fits and the cubic splines match the
DSE data almost perfectly and yield very similar results. In fact, this
observation holds true for real chemical potentials up to rather large values.
As a consequence and for the sake of comparability with the lattice results, we
restrict ourselves to the polynomial fit results except when noted otherwise.

From Fig.~\ref{fig:immu_fits} it is obvious that our curves $\chi(\cond)$ at
different rescaled imaginary chemical potential $\mu_B/T$ are close to each
other, but do not collapse to one curve. We rather find that both peak height
and peak position vary slightly with imaginary $\mu_B$. That is, the peak
position is at the largest $\chi$ but smallest $\cond$ for $\muB / T = 0$ while
it moves to smallest $\chi$ but largest $\cond$ for $\muB / T = 7/8 \pi \ii$. We
therefore do not confirm the apparent collapse (within error bars) observed in
Ref.~\cite{Borsanyi:2020fev}, however without attaching great significance to
this observation. Indeed, in later works, $\mu_B-$dependent corrections to
simple rescaling laws have been successfully explored
\cite{Borsanyi:2021sxv,Borsanyi2022a}.

In the next step, we gauge the quality of extrapolations from imaginary to real
chemical potentials. In the DSE approach, this is possible since explicit
results at real chemical potential have been determined already in
Ref.~\cite{Gunkel:2021oya} and have been verified by us independently for this
work. Using first the inflection point of the subtracted condensate as crossover
criterion, the phase diagram for the setup of Ref.~\cite{Gunkel:2021oya}
(``previous work'') and the one of this work with the slightly adapted vertex
strength parameter $d_1$ (``inflection point'') are shown in
Fig.~\ref{fig:both_Tc}. We find an almost uniformly shifted phase boundary to
slightly lower temperatures as expected due to the reduced vertex strength.
Using the peak of the susceptibility, Eq.~\eqref{cond}, as criterion, we find a
crossover line at somewhat larger temperatures that merges with the previous
one, of course, at the critical endpoint. The fits to the condensate using
polynomials as well as the cubic spline interpolations end up in very similar
results.

Now we are in a position to compare results obtained from extrapolations with
explicitly calculated results. To this end, we fit our results for the phase
boundary at imaginary chemical potentials to the well-known parametrization of
the pseudocritical transition temperature:
\begin{equation}
    \label{eq:parametrization_Tc}
    \frac{\Tc(\muB)}{\Tc}
    =
    1
    -
    \kappa_{2} \biggl( \frac{\muB}{\Tc} \biggr)^2
    -
    \kappa_{4} \biggl( \frac{\muB}{\Tc} \biggr)^4
    \,,
\end{equation}
where $\Tc = \Tc(\muB = 0)$ and $\kappa_{2}$ labels the curvature. In the left
half of the lower diagram of Fig.~\ref{fig:both_Tc}, we illustrate the
pseudocritical temperature obtained for polynomial fits at imaginary chemical
potentials. The blue square boxes show the results of the DSE calculations,
while the black curve indicates the fit to the parametrization given in
Eq.~\eqref{eq:parametrization_Tc}. The resulting fit parameters are given by:
\begin{equation}\label{kappa}
    \kappa_{2}^{\text{poly}} = 0.0196 \pm 0.0001
    \,,
    \quad
    \kappa_{4}^{\text{poly}} = 0.00015 \pm 0.00001
    \,,
\end{equation}
with very small error and almost vanishing $\chi^2$-value due to the non-statistical 
nature of the data. The fit very accurately
represents the DSE data. If we use cubic spline interpolations, we again end up
at very similar results, i.e., $\kappa_{2}^{\text{spln}} = 0.0196 \pm 0.0001$ and
$\kappa_{4}^{\text{spln}} = 0.00014 \pm 0.00001$. This underpins the fact that at imaginary
chemical potentials, our data points are very well represented by the polynomial
fit. The difference of the values for $\kappa_{2}$ in Eq.~\eqref{kappa} as
compared to
\begin{equation}
    \kappa_{2}^{\text{infl}}
    =
    0.0173
\end{equation}
presented in Ref.~\cite{Gunkel:2021oya} and obtained in the same truncation scheme 
as used in this work is entirely due to the difference in the definition of the 
pseudocritical temperature from a different regularized order parameter. The 
resulting shifts of the transition lines seen in the upper diagram of Fig.~\ref{fig:both_Tc} 
has been discussed above. Since pseudo-critical transition lines from different 
definitions all end up in the same critical endpoint they naturally have different 
curvature. For completeness, we show results from different sources in Fig.~\ref{fig:kappa}
and the resulting critical end point for the DSE/FRG calculations in the upper right 
diagram of Fig.~\ref{fig:both_Tc}. 

Now, we can analytically continue this fit to real chemical potential and
compare to DSE results obtained there. This is done and displayed in the middle 
diagram of Fig.~\ref{fig:both_Tc}. Starting with fits to data at zero chemical potential
and the two lowest imaginary chemical potential values $N=3$, we add point by 
point 
in the direction of large imaginary chemical potential until we included all data at $N=8$.
We clearly observe that in the quality of the fits improve rapidly until it settles down 
at about $N=6$. Adding further points brings only very small improvement. We conclude
from this that going deep in the imaginary chemical potential region is important, but
one does not need to go all the way to the Roberge-Weiss bound. We could have stopped at
$N=6$. We furthermore checked that leaving out two arbitrary points in between 
the one
with largest imaginary chemical potential and the one with zero chemical potential does
not change the quality of the fit at all. 

In the lower diagram of Fig.~\ref{fig:both_Tc}, we show our best fit result 
together 
with the lattice data of Ref.~\cite{Borsanyi:2020fev} (red diamonds) and their fit result (grey band).
The discrepancy of our data and the lattice data at large imaginary chemical potential 
is a measure of the systematic error of our calculation and therefore of our data points.
This systematic error is small by construction in the vicinity of zero chemical potential,
but grows into the five percent region for largest imaginary chemical potential. (Naively) 
assuming the error to be symmetric in $\mu_B$, our CEP could be accurate on the level of 
ten percent. But, of course, we cannot exclude that new physics effects at large real chemical 
potential might invalidate this naive estimate. 

In general, we observe a remarkable coincidence between our fit and the DSE data up to rather large
chemical potentials. Deviations of more than one percent in temperature occur
only for $\muB > \SI{510}{\MeV}$, i.e. for chemical potential larger than 80 \%
of the one of the CEP, which in the present truncation is located at around
$\muB^{\textup{CEP}} \approx \SI{636}{\MeV}$.
We can infer that the extrapolation from imaginary chemical potentials works
excellently for a large part of the crossover region well towards the CEP.

In vicinity of the CEP, however, not only does the extrapolation of $\Tc$ cease
to be valid, the $\chi(\cond)$ curves also can no longer be approximated by a
polynomial as is illustrated in Fig.~\ref{fig:remu_fits}. There, we show the
chiral susceptibility as a function of the chiral condensate analogously to
Fig.~\ref{fig:immu_fits}, this time for real chemical potentials $\muB
>\SI{420}{\MeV}$ and $\muB = 0$ as a reference. We depict the polynomial fits
and cubic spline interpolations in the left and right panels, respectively.
Obviously, the approximation of a fifth-order polynomial breaks down completely
at around $\muB \gtrsim \SI{600}{\MeV}$. In this region, only the splines
reliably coincide with the DSE data.

This discrepancy is also visible in Fig.~\ref{fig:both_reg}, where we plot peak
height and peak position of $\chi(\cond)$ as a function of negative and positive
values of $(\muB / T)^{2}$ in Fig.~\ref{fig:both_reg}. We show results for the
polynomial fits as well as the cubic spline interpolations for the sake of
comparison. We notice again that both peak position and height are not entirely
constant at imaginary chemical potentials in slight contrast to the lattice
results discussed in Ref.~\cite{Borsanyi:2020fev}. While the change of the
condensate is more or less constant for real $\muB$, the critical susceptibility
does not change much for small chemical potentials but drastically in vicinity
of the CEP as expected. This is also the region where the polynomial fit and the
spline interpolation deviate substantially, i.e., where the assumption of a low
order polynomial breaks down. Again, this is to be expected since the
susceptibility is singular at the CEP.

We would like to finish our analysis by pointing out that our results fully support 
the analysis performed in Ref.~\cite{Borsanyi:2020fev}. In the imaginary chemical
potential region where the lattice fits have been done, there is no difference 
in quality 
in using polynomial or spline fits. This is also true for the region of real chemical 
potential covered by the lattice extrapolation. The (expected) problems with the 
polynomial fits emerge only at much larger real chemical potential, close to the CEP. 

\section{\label{summary}%
    Summary and conclusions
}

In this work, we have studied whether the extrapolation procedure from imaginary
to real chemical potential introduced in Ref.~\cite{Borsanyi:2020fev} in the
context of lattice QCD is capable to reproduce explicit results for the phase
transition line at real chemical potential obtained with functional methods. The
result is very encouraging: up to quite large chemical potentials not very much
smaller ($\approx 20 \%$) than the one of the critical endpoint, the
extrapolation works extremely well. For larger chemical potential, the
extrapolated transition line undershoots the calculated one; at the critical
chemical potential $\muB^{\textup{CEP}} \approx \SI{636}{\MeV}$, the resulting
temperature of the extrapolation is about $\SI{13}{\MeV}$ too small. Also, of
course, it is not possible to extract the location of the CEP from the
extrapolation procedure. However, comparing the lattice results with 
the ones of our framework at large imaginary chemical potential and taking this
as a (naive) measure for the corresponding systematic error of our framework at
corresponding values for real chemical potential, this indicates a systematic error 
of the order of 5-10 percent for the location of the CEP. 

An interesting option is to study the different extrapolation procedure 
brought forward in \cite{Dimopoulos:2021vrk}. An interesting further option is
to directly determine the location of Lee-Young zeros. This is relegated to future work.


\begin{acknowledgments}
We thank both, Jana N.~Guenther and Philipp Isserstedt for enlightening
discussions. We furthermore thank Jana N.~Guenther for providing the lattice
data for comparison and Philipp Isserstedt for crosschecks of the numerical
code at an early stage of this work. This work has been supported by the
Helmholtz Graduate School for Hadron and Ion Research (HGS-HIRe) for FAIR, the
GSI Helmholtzzentrum f\"{u}r Schwerionenforschung and the Deutsche
Forschungsgemeinschaft (DFG, German Research Foundation) through the
Collaborative Research Center TransRegio CRC-TR 211 ``Strong-interaction matter
under extreme conditions''.\\
Feynman diagrams were drawn with \textsc{JaxoDraw} \cite{Binosi:2008ig}.
\end{acknowledgments}


\bibliography{ImaginaryMuBibliography}

\begin{thebibliography}{43}%
\makeatletter
\providecommand \@ifxundefined [1]{%
 \@ifx{#1\undefined}
}%
\providecommand \@ifnum [1]{%
 \ifnum #1\expandafter \@firstoftwo
 \else \expandafter \@secondoftwo
 \fi
}%
\providecommand \@ifx [1]{%
 \ifx #1\expandafter \@firstoftwo
 \else \expandafter \@secondoftwo
 \fi
}%
\providecommand \natexlab [1]{#1}%
\providecommand \enquote  [1]{``#1''}%
\providecommand \bibnamefont  [1]{#1}%
\providecommand \bibfnamefont [1]{#1}%
\providecommand \citenamefont [1]{#1}%
\providecommand \href@noop [0]{\@secondoftwo}%
\providecommand \href [0]{\begingroup \@sanitize@url \@href}%
\providecommand \@href[1]{\@@startlink{#1}\@@href}%
\providecommand \@@href[1]{\endgroup#1\@@endlink}%
\providecommand \@sanitize@url [0]{\catcode `\\12\catcode `\$12\catcode
  `\&12\catcode `\#12\catcode `\^12\catcode `\_12\catcode `\%12\relax}%
\providecommand \@@startlink[1]{}%
\providecommand \@@endlink[0]{}%
\providecommand \url  [0]{\begingroup\@sanitize@url \@url }%
\providecommand \@url [1]{\endgroup\@href {#1}{\urlprefix }}%
\providecommand \urlprefix  [0]{URL }%
\providecommand \Eprint [0]{\href }%
\providecommand \doibase [0]{https://doi.org/}%
\providecommand \selectlanguage [0]{\@gobble}%
\providecommand \bibinfo  [0]{\@secondoftwo}%
\providecommand \bibfield  [0]{\@secondoftwo}%
\providecommand \translation [1]{[#1]}%
\providecommand \BibitemOpen [0]{}%
\providecommand \bibitemStop [0]{}%
\providecommand \bibitemNoStop [0]{.\EOS\space}%
\providecommand \EOS [0]{\spacefactor3000\relax}%
\providecommand \BibitemShut  [1]{\csname bibitem#1\endcsname}%
\let\auto@bib@innerbib\@empty
\bibitem [{\citenamefont {Bors\'{a}nyi}\ \emph {et~al.}(2020)\citenamefont
  {Bors\'{a}nyi} \emph {et~al.}}]{Borsanyi:2020fev}%
  \BibitemOpen
  \bibfield  {author} {\bibinfo {author} {\bibfnamefont {S.}~\bibnamefont
  {Bors\'{a}nyi}} \emph {et~al.},\ }\bibfield  {title} {\bibinfo {title} {{QCD
  Crossover at Finite Chemical Potential from Lattice Simulations}},\ }\href
  {https://doi.org/10.1103/PhysRevLett.125.052001} {\bibfield  {journal}
  {\bibinfo  {journal} {Phys. Rev. Lett.}\ }\textbf {\bibinfo {volume} {125}},\
  \bibinfo {pages} {052001} (\bibinfo {year} {2020})},\ \Eprint
  {https://arxiv.org/abs/2002.02821} {arXiv:2002.02821 [hep-lat]} \BibitemShut
  {NoStop}%
\bibitem [{\citenamefont {Bzdak}\ \emph {et~al.}(2020)\citenamefont {Bzdak}
  \emph {et~al.}}]{Bzdak:2019pkr}%
  \BibitemOpen
  \bibfield  {author} {\bibinfo {author} {\bibfnamefont {A.}~\bibnamefont
  {Bzdak}} \emph {et~al.},\ }\bibfield  {title} {\bibinfo {title} {{Mapping the
  phases of quantum chromodynamics with beam energy scan}},\ }\href
  {https://doi.org/10.1016/j.physrep.2020.01.005} {\bibfield  {journal}
  {\bibinfo  {journal} {Phys. Rep.}\ }\textbf {\bibinfo {volume} {853}},\
  \bibinfo {pages} {1} (\bibinfo {year} {2020})},\ \Eprint
  {https://arxiv.org/abs/1906.00936} {arXiv:1906.00936 [nucl-th]} \BibitemShut
  {NoStop}%
\bibitem [{\citenamefont {Salabura}\ and\ \citenamefont
  {Stroth}(2021)}]{Salabura:2020tou}%
  \BibitemOpen
  \bibfield  {author} {\bibinfo {author} {\bibfnamefont {P.}~\bibnamefont
  {Salabura}}\ and\ \bibinfo {author} {\bibfnamefont {J.}~\bibnamefont
  {Stroth}},\ }\bibfield  {title} {\bibinfo {title} {{Dilepton radiation from
  strongly interacting systems}},\ }\href
  {https://doi.org/10.1016/j.ppnp.2021.103869} {\bibfield  {journal} {\bibinfo
  {journal} {Prog. Part. Nucl. Phys.}\ }\textbf {\bibinfo {volume} {120}},\
  \bibinfo {pages} {103869} (\bibinfo {year} {2021})},\ \Eprint
  {https://arxiv.org/abs/2005.14589} {arXiv:2005.14589 [nucl-ex]} \BibitemShut
  {NoStop}%
\bibitem [{\citenamefont {Friman}\ \emph {et~al.}(2011)\citenamefont {Friman}
  \emph {et~al.}}]{Friman:2011zz}%
  \BibitemOpen
  \bibinfo {editor} {\bibfnamefont {B.}~\bibnamefont {Friman}} \emph {et~al.},\
  eds.,\ \href {https://doi.org/10.1007/978-3-642-13293-3} {\emph {\bibinfo
  {title} {{The CBM Physics Book: Compressed Baryonic Matter in Laboratory
  Experiments}}}},\ \bibinfo {series} {{Lecture Notes in Physics}}\ No.\
  \bibinfo {number} {814}\ (\bibinfo  {publisher} {Springer},\ \bibinfo {year}
  {2011})\BibitemShut {NoStop}%
\bibitem [{\citenamefont {Nagata}(2022)}]{Nagata2022}%
  \BibitemOpen
  \bibfield  {author} {\bibinfo {author} {\bibfnamefont {K.}~\bibnamefont
  {Nagata}},\ }\bibfield  {title} {\bibinfo {title} {{Finite-density lattice
  QCD and sign problem: Current status and open problems}},\ }\href
  {https://doi.org/10.1016/j.ppnp.2022.103991} {\bibfield  {journal} {\bibinfo
  {journal} {Prog. Part. Nucl. Phys.}\ }\textbf {\bibinfo {volume} {127}},\
  \bibinfo {pages} {103991} (\bibinfo {year} {2022})},\ \Eprint
  {https://arxiv.org/abs/2108.12423} {arXiv:2108.12423 [hep-lat]} \BibitemShut
  {NoStop}%
\bibitem [{\citenamefont {Hasenfratz}\ and\ \citenamefont
  {Toussaint}(1992)}]{Hasenfratz:1991ax}%
  \BibitemOpen
  \bibfield  {author} {\bibinfo {author} {\bibfnamefont {A.}~\bibnamefont
  {Hasenfratz}}\ and\ \bibinfo {author} {\bibfnamefont {D.}~\bibnamefont
  {Toussaint}},\ }\bibfield  {title} {\bibinfo {title} {{Canonical ensembles
  and nonzero density quantum chromodynamics}},\ }\href
  {https://doi.org/10.1016/0550-3213(92)90247-9} {\bibfield  {journal}
  {\bibinfo  {journal} {Nucl. Phys. B}\ }\textbf {\bibinfo {volume} {371}},\
  \bibinfo {pages} {539} (\bibinfo {year} {1992})}\BibitemShut {NoStop}%
\bibitem [{\citenamefont {Fodor}\ and\ \citenamefont
  {Katz}(2002{\natexlab{a}})}]{Fodor:2001au}%
  \BibitemOpen
  \bibfield  {author} {\bibinfo {author} {\bibfnamefont {Z.}~\bibnamefont
  {Fodor}}\ and\ \bibinfo {author} {\bibfnamefont {S.~D.}\ \bibnamefont
  {Katz}},\ }\bibfield  {title} {\bibinfo {title} {{A New method to study
  lattice QCD at finite temperature and chemical potential}},\ }\href
  {https://doi.org/10.1016/S0370-2693(02)01583-6} {\bibfield  {journal}
  {\bibinfo  {journal} {Phys. Lett. B}\ }\textbf {\bibinfo {volume} {534}},\
  \bibinfo {pages} {87} (\bibinfo {year} {2002}{\natexlab{a}})},\ \Eprint
  {https://arxiv.org/abs/hep-lat/0104001} {arXiv:hep-lat/0104001} \BibitemShut
  {NoStop}%
\bibitem [{\citenamefont {Fodor}\ and\ \citenamefont
  {Katz}(2002{\natexlab{b}})}]{Fodor:2001pe}%
  \BibitemOpen
  \bibfield  {author} {\bibinfo {author} {\bibfnamefont {Z.}~\bibnamefont
  {Fodor}}\ and\ \bibinfo {author} {\bibfnamefont {S.~D.}\ \bibnamefont
  {Katz}},\ }\bibfield  {title} {\bibinfo {title} {{Lattice determination of
  the critical point of QCD at finite T and mu}},\ }\href
  {https://doi.org/10.1088/1126-6708/2002/03/014} {\bibfield  {journal}
  {\bibinfo  {journal} {JHEP}\ }\textbf {\bibinfo {volume} {03}},\ \bibinfo
  {pages} {014}},\ \Eprint {https://arxiv.org/abs/hep-lat/0106002}
  {arXiv:hep-lat/0106002} \BibitemShut {NoStop}%
\bibitem [{\citenamefont {Giordano}\ \emph
  {et~al.}(2020{\natexlab{a}})\citenamefont {Giordano}, \citenamefont {Kapas},
  \citenamefont {Katz}, \citenamefont {Nogradi},\ and\ \citenamefont
  {Pasztor}}]{Giordano:2020roi}%
  \BibitemOpen
  \bibfield  {author} {\bibinfo {author} {\bibfnamefont {M.}~\bibnamefont
  {Giordano}}, \bibinfo {author} {\bibfnamefont {K.}~\bibnamefont {Kapas}},
  \bibinfo {author} {\bibfnamefont {S.~D.}\ \bibnamefont {Katz}}, \bibinfo
  {author} {\bibfnamefont {D.}~\bibnamefont {Nogradi}},\ and\ \bibinfo {author}
  {\bibfnamefont {A.}~\bibnamefont {Pasztor}},\ }\bibfield  {title} {\bibinfo
  {title} {{New approach to lattice QCD at finite density; results for the
  critical end point on coarse lattices}},\ }\href
  {https://doi.org/10.1007/JHEP05(2020)088} {\bibfield  {journal} {\bibinfo
  {journal} {JHEP}\ }\textbf {\bibinfo {volume} {05}},\ \bibinfo {pages}
  {088}},\ \Eprint {https://arxiv.org/abs/2004.10800} {arXiv:2004.10800
  [hep-lat]} \BibitemShut {NoStop}%
\bibitem [{\citenamefont {Allton}\ \emph {et~al.}(2002)\citenamefont {Allton},
  \citenamefont {Ejiri}, \citenamefont {Hands}, \citenamefont {Kaczmarek},
  \citenamefont {Karsch}, \citenamefont {Laermann}, \citenamefont {Schmidt},\
  and\ \citenamefont {Scorzato}}]{Allton:2002zi}%
  \BibitemOpen
  \bibfield  {author} {\bibinfo {author} {\bibfnamefont {C.~R.}\ \bibnamefont
  {Allton}}, \bibinfo {author} {\bibfnamefont {S.}~\bibnamefont {Ejiri}},
  \bibinfo {author} {\bibfnamefont {S.~J.}\ \bibnamefont {Hands}}, \bibinfo
  {author} {\bibfnamefont {O.}~\bibnamefont {Kaczmarek}}, \bibinfo {author}
  {\bibfnamefont {F.}~\bibnamefont {Karsch}}, \bibinfo {author} {\bibfnamefont
  {E.}~\bibnamefont {Laermann}}, \bibinfo {author} {\bibfnamefont
  {C.}~\bibnamefont {Schmidt}},\ and\ \bibinfo {author} {\bibfnamefont
  {L.}~\bibnamefont {Scorzato}},\ }\bibfield  {title} {\bibinfo {title} {{The
  QCD thermal phase transition in the presence of a small chemical
  potential}},\ }\href {https://doi.org/10.1103/PhysRevD.66.074507} {\bibfield
  {journal} {\bibinfo  {journal} {Phys. Rev. D}\ }\textbf {\bibinfo {volume}
  {66}},\ \bibinfo {pages} {074507} (\bibinfo {year} {2002})},\ \Eprint
  {https://arxiv.org/abs/hep-lat/0204010} {arXiv:hep-lat/0204010} \BibitemShut
  {NoStop}%
\bibitem [{\citenamefont {Gavai}\ and\ \citenamefont
  {Gupta}(2008)}]{Gavai:2008zr}%
  \BibitemOpen
  \bibfield  {author} {\bibinfo {author} {\bibfnamefont {R.~V.}\ \bibnamefont
  {Gavai}}\ and\ \bibinfo {author} {\bibfnamefont {S.}~\bibnamefont {Gupta}},\
  }\bibfield  {title} {\bibinfo {title} {{QCD at finite chemical potential with
  six time slices}},\ }\href {https://doi.org/10.1103/PhysRevD.78.114503}
  {\bibfield  {journal} {\bibinfo  {journal} {Phys. Rev. D}\ }\textbf {\bibinfo
  {volume} {78}},\ \bibinfo {pages} {114503} (\bibinfo {year} {2008})},\
  \Eprint {https://arxiv.org/abs/0806.2233} {arXiv:0806.2233 [hep-lat]}
  \BibitemShut {NoStop}%
\bibitem [{\citenamefont {Borsanyi}\ \emph {et~al.}(2012)\citenamefont
  {Borsanyi}, \citenamefont {Endrodi}, \citenamefont {Fodor}, \citenamefont
  {Katz}, \citenamefont {Krieg}, \citenamefont {Ratti},\ and\ \citenamefont
  {Szabo}}]{Borsanyi:2012cr}%
  \BibitemOpen
  \bibfield  {author} {\bibinfo {author} {\bibfnamefont {S.}~\bibnamefont
  {Borsanyi}}, \bibinfo {author} {\bibfnamefont {G.}~\bibnamefont {Endrodi}},
  \bibinfo {author} {\bibfnamefont {Z.}~\bibnamefont {Fodor}}, \bibinfo
  {author} {\bibfnamefont {S.~D.}\ \bibnamefont {Katz}}, \bibinfo {author}
  {\bibfnamefont {S.}~\bibnamefont {Krieg}}, \bibinfo {author} {\bibfnamefont
  {C.}~\bibnamefont {Ratti}},\ and\ \bibinfo {author} {\bibfnamefont {K.~K.}\
  \bibnamefont {Szabo}},\ }\bibfield  {title} {\bibinfo {title} {{QCD equation
  of state at nonzero chemical potential: continuum results with physical quark
  masses at order $mu^2$}},\ }\href {https://doi.org/10.1007/JHEP08(2012)053}
  {\bibfield  {journal} {\bibinfo  {journal} {JHEP}\ }\textbf {\bibinfo
  {volume} {08}},\ \bibinfo {pages} {053}},\ \Eprint
  {https://arxiv.org/abs/1204.6710} {arXiv:1204.6710 [hep-lat]} \BibitemShut
  {NoStop}%
\bibitem [{\citenamefont {Bazavov}\ \emph {et~al.}(2017)\citenamefont {Bazavov}
  \emph {et~al.}}]{Bazavov:2017dus}%
  \BibitemOpen
  \bibfield  {author} {\bibinfo {author} {\bibfnamefont {A.}~\bibnamefont
  {Bazavov}} \emph {et~al.},\ }\bibfield  {title} {\bibinfo {title} {{The QCD
  Equation of State to $\mathcal{O}(\mu_B^6)$ from Lattice QCD}},\ }\href
  {https://doi.org/10.1103/PhysRevD.95.054504} {\bibfield  {journal} {\bibinfo
  {journal} {Phys. Rev. D}\ }\textbf {\bibinfo {volume} {95}},\ \bibinfo
  {pages} {054504} (\bibinfo {year} {2017})},\ \Eprint
  {https://arxiv.org/abs/1701.04325} {arXiv:1701.04325 [hep-lat]} \BibitemShut
  {NoStop}%
\bibitem [{\citenamefont {Giordano}\ \emph
  {et~al.}(2020{\natexlab{b}})\citenamefont {Giordano}, \citenamefont {Kapas},
  \citenamefont {Katz}, \citenamefont {Nogradi},\ and\ \citenamefont
  {Pasztor}}]{Giordano:2019gev}%
  \BibitemOpen
  \bibfield  {author} {\bibinfo {author} {\bibfnamefont {M.}~\bibnamefont
  {Giordano}}, \bibinfo {author} {\bibfnamefont {K.}~\bibnamefont {Kapas}},
  \bibinfo {author} {\bibfnamefont {S.~D.}\ \bibnamefont {Katz}}, \bibinfo
  {author} {\bibfnamefont {D.}~\bibnamefont {Nogradi}},\ and\ \bibinfo {author}
  {\bibfnamefont {A.}~\bibnamefont {Pasztor}},\ }\bibfield  {title} {\bibinfo
  {title} {{Radius of convergence in lattice QCD at finite $\mu_B$ with rooted
  staggered fermions}},\ }\href {https://doi.org/10.1103/PhysRevD.101.074511}
  {\bibfield  {journal} {\bibinfo  {journal} {Phys. Rev. D}\ }\textbf {\bibinfo
  {volume} {101}},\ \bibinfo {pages} {074511} (\bibinfo {year}
  {2020}{\natexlab{b}})},\ \bibinfo {note} {[Erratum: Phys.Rev.D 104, 119901
  (2021)]},\ \Eprint {https://arxiv.org/abs/1911.00043} {arXiv:1911.00043
  [hep-lat]} \BibitemShut {NoStop}%
\bibitem [{\citenamefont {Bazavov}\ \emph
  {et~al.}(2019{\natexlab{a}})\citenamefont {Bazavov} \emph
  {et~al.}}]{Bazavov2019}%
  \BibitemOpen
  \bibfield  {author} {\bibinfo {author} {\bibfnamefont {A.}~\bibnamefont
  {Bazavov}} \emph {et~al.} (\bibinfo {collaboration} {HotQCD}),\ }\bibfield
  {title} {\bibinfo {title} {{Chiral crossover in QCD at zero and non-zero
  chemical potentials}},\ }\href
  {https://doi.org/10.1016/j.physletb.2019.05.013} {\bibfield  {journal}
  {\bibinfo  {journal} {Phys. Lett. B}\ }\textbf {\bibinfo {volume} {795}},\
  \bibinfo {pages} {15} (\bibinfo {year} {2019}{\natexlab{a}})},\ \Eprint
  {https://arxiv.org/abs/1812.08235} {arXiv:1812.08235 [hep-lat]} \BibitemShut
  {NoStop}%
\bibitem [{\citenamefont {Bazavov}\ \emph {et~al.}(2020)\citenamefont {Bazavov}
  \emph {et~al.}}]{Bazavov:2020bjn}%
  \BibitemOpen
  \bibfield  {author} {\bibinfo {author} {\bibfnamefont {A.}~\bibnamefont
  {Bazavov}} \emph {et~al.},\ }\bibfield  {title} {\bibinfo {title} {{Skewness,
  kurtosis, and the fifth and sixth order cumulants of net baryon-number
  distributions from lattice QCD confront high-statistics STAR data}},\ }\href
  {https://doi.org/10.1103/PhysRevD.101.074502} {\bibfield  {journal} {\bibinfo
   {journal} {Phys. Rev. D}\ }\textbf {\bibinfo {volume} {101}},\ \bibinfo
  {pages} {074502} (\bibinfo {year} {2020})},\ \Eprint
  {https://arxiv.org/abs/2001.08530} {arXiv:2001.08530 [hep-lat]} \BibitemShut
  {NoStop}%
\bibitem [{\citenamefont {Bollweg}\ \emph {et~al.}(2022)\citenamefont
  {Bollweg}, \citenamefont {Goswami}, \citenamefont {Kaczmarek}, \citenamefont
  {Karsch}, \citenamefont {Mukherjee}, \citenamefont {Petreczky}, \citenamefont
  {Schmidt},\ and\ \citenamefont {Scior}}]{Bollweg2022}%
  \BibitemOpen
  \bibfield  {author} {\bibinfo {author} {\bibfnamefont {D.}~\bibnamefont
  {Bollweg}}, \bibinfo {author} {\bibfnamefont {J.}~\bibnamefont {Goswami}},
  \bibinfo {author} {\bibfnamefont {O.}~\bibnamefont {Kaczmarek}}, \bibinfo
  {author} {\bibfnamefont {F.}~\bibnamefont {Karsch}}, \bibinfo {author}
  {\bibfnamefont {S.}~\bibnamefont {Mukherjee}}, \bibinfo {author}
  {\bibfnamefont {P.}~\bibnamefont {Petreczky}}, \bibinfo {author}
  {\bibfnamefont {C.}~\bibnamefont {Schmidt}},\ and\ \bibinfo {author}
  {\bibfnamefont {P.}~\bibnamefont {Scior}} (\bibinfo {collaboration}
  {HotQCD}),\ }\bibfield  {title} {\bibinfo {title} {{Taylor expansions and
  Pad\'e approximants for cumulants of conserved charge fluctuations at
  nonvanishing chemical potentials}},\ }\href
  {https://doi.org/10.1103/PhysRevD.105.074511} {\bibfield  {journal} {\bibinfo
   {journal} {Phys. Rev. D}\ }\textbf {\bibinfo {volume} {105}},\ \bibinfo
  {pages} {074511} (\bibinfo {year} {2022})},\ \Eprint
  {https://arxiv.org/abs/2202.09184} {arXiv:2202.09184 [hep-lat]} \BibitemShut
  {NoStop}%
\bibitem [{\citenamefont {Ratti}(2023)}]{Ratti:2022qgf}%
  \BibitemOpen
  \bibfield  {author} {\bibinfo {author} {\bibfnamefont {C.}~\bibnamefont
  {Ratti}},\ }\bibfield  {title} {\bibinfo {title} {{Equation of state for QCD
  from lattice simulations}},\ }\href
  {https://doi.org/10.1016/j.ppnp.2022.104007} {\bibfield  {journal} {\bibinfo
  {journal} {Prog. Part. Nucl. Phys.}\ }\textbf {\bibinfo {volume} {129}},\
  \bibinfo {pages} {104007} (\bibinfo {year} {2023})}\BibitemShut {NoStop}%
\bibitem [{\citenamefont {de~Forcrand}\ and\ \citenamefont
  {Philipsen}(2002)}]{deForcrand:2002hgr}%
  \BibitemOpen
  \bibfield  {author} {\bibinfo {author} {\bibfnamefont {P.}~\bibnamefont
  {de~Forcrand}}\ and\ \bibinfo {author} {\bibfnamefont {O.}~\bibnamefont
  {Philipsen}},\ }\bibfield  {title} {\bibinfo {title} {{The QCD phase diagram
  for small densities from imaginary chemical potential}},\ }\href
  {https://doi.org/10.1016/S0550-3213(02)00626-0} {\bibfield  {journal}
  {\bibinfo  {journal} {Nucl. Phys. B}\ }\textbf {\bibinfo {volume} {642}},\
  \bibinfo {pages} {290} (\bibinfo {year} {2002})},\ \Eprint
  {https://arxiv.org/abs/hep-lat/0205016} {arXiv:hep-lat/0205016} \BibitemShut
  {NoStop}%
\bibitem [{\citenamefont {Bors\'{a}nyi}\ \emph {et~al.}(2021)\citenamefont
  {Bors\'{a}nyi} \emph {et~al.}}]{Borsanyi:2021sxv}%
  \BibitemOpen
  \bibfield  {author} {\bibinfo {author} {\bibfnamefont {S.}~\bibnamefont
  {Bors\'{a}nyi}} \emph {et~al.},\ }\bibfield  {title} {\bibinfo {title}
  {{Lattice QCD Equation of State at Finite Chemical Potential from an
  Alternative Expansion Scheme}},\ }\href
  {https://doi.org/10.1103/PhysRevLett.126.232001} {\bibfield  {journal}
  {\bibinfo  {journal} {Phys. Rev. Lett.}\ }\textbf {\bibinfo {volume} {126}},\
  \bibinfo {pages} {232001} (\bibinfo {year} {2021})},\ \Eprint
  {https://arxiv.org/abs/2102.06660} {arXiv:2102.06660 [hep-lat]} \BibitemShut
  {NoStop}%
\bibitem [{\citenamefont {Dimopoulos}\ \emph {et~al.}(2022)\citenamefont
  {Dimopoulos}, \citenamefont {Dini}, \citenamefont {Di~Renzo}, \citenamefont
  {Goswami}, \citenamefont {Nicotra}, \citenamefont {Schmidt}, \citenamefont
  {Singh}, \citenamefont {Zambello},\ and\ \citenamefont
  {Ziesch\'e}}]{Dimopoulos:2021vrk}%
  \BibitemOpen
  \bibfield  {author} {\bibinfo {author} {\bibfnamefont {P.}~\bibnamefont
  {Dimopoulos}}, \bibinfo {author} {\bibfnamefont {L.}~\bibnamefont {Dini}},
  \bibinfo {author} {\bibfnamefont {F.}~\bibnamefont {Di~Renzo}}, \bibinfo
  {author} {\bibfnamefont {J.}~\bibnamefont {Goswami}}, \bibinfo {author}
  {\bibfnamefont {G.}~\bibnamefont {Nicotra}}, \bibinfo {author} {\bibfnamefont
  {C.}~\bibnamefont {Schmidt}}, \bibinfo {author} {\bibfnamefont
  {S.}~\bibnamefont {Singh}}, \bibinfo {author} {\bibfnamefont
  {K.}~\bibnamefont {Zambello}},\ and\ \bibinfo {author} {\bibfnamefont
  {F.}~\bibnamefont {Ziesch\'e}},\ }\bibfield  {title} {\bibinfo {title}
  {{Contribution to understanding the phase structure of strong interaction
  matter: Lee-Yang edge singularities from lattice QCD}},\ }\href
  {https://doi.org/10.1103/PhysRevD.105.034513} {\bibfield  {journal} {\bibinfo
   {journal} {Phys. Rev. D}\ }\textbf {\bibinfo {volume} {105}},\ \bibinfo
  {pages} {034513} (\bibinfo {year} {2022})},\ \Eprint
  {https://arxiv.org/abs/2110.15933} {arXiv:2110.15933 [hep-lat]} \BibitemShut
  {NoStop}%
\bibitem [{\citenamefont {Borsanyi}\ \emph
  {et~al.}(2022{\natexlab{a}})\citenamefont {Borsanyi}, \citenamefont {Fodor},
  \citenamefont {Giordano}, \citenamefont {Guenther}, \citenamefont {Katz},
  \citenamefont {Pasztor},\ and\ \citenamefont {Wong}}]{Borsanyi:2022soo}%
  \BibitemOpen
  \bibfield  {author} {\bibinfo {author} {\bibfnamefont {S.}~\bibnamefont
  {Borsanyi}}, \bibinfo {author} {\bibfnamefont {Z.}~\bibnamefont {Fodor}},
  \bibinfo {author} {\bibfnamefont {M.}~\bibnamefont {Giordano}}, \bibinfo
  {author} {\bibfnamefont {J.~N.}\ \bibnamefont {Guenther}}, \bibinfo {author}
  {\bibfnamefont {S.~D.}\ \bibnamefont {Katz}}, \bibinfo {author}
  {\bibfnamefont {A.}~\bibnamefont {Pasztor}},\ and\ \bibinfo {author}
  {\bibfnamefont {C.~H.}\ \bibnamefont {Wong}},\ }\bibfield  {title} {\bibinfo
  {title} {{Equation of state of a hot-and-dense quark gluon plasma: lattice
  simulations at real $\mu_B$ vs. extrapolations}},\ }\href@noop {} {\
  (\bibinfo {year} {2022}{\natexlab{a}})},\ \Eprint
  {https://arxiv.org/abs/2208.05398} {arXiv:2208.05398 [hep-lat]} \BibitemShut
  {NoStop}%
\bibitem [{\citenamefont {Borsanyi}\ \emph
  {et~al.}(2022{\natexlab{b}})\citenamefont {Borsanyi}, \citenamefont
  {Guenther}, \citenamefont {Kara}, \citenamefont {Fodor}, \citenamefont
  {Parotto}, \citenamefont {Pasztor}, \citenamefont {Ratti},\ and\
  \citenamefont {Szabo}}]{Borsanyi2022a}%
  \BibitemOpen
  \bibfield  {author} {\bibinfo {author} {\bibfnamefont {S.}~\bibnamefont
  {Borsanyi}}, \bibinfo {author} {\bibfnamefont {J.~N.}\ \bibnamefont
  {Guenther}}, \bibinfo {author} {\bibfnamefont {R.}~\bibnamefont {Kara}},
  \bibinfo {author} {\bibfnamefont {Z.}~\bibnamefont {Fodor}}, \bibinfo
  {author} {\bibfnamefont {P.}~\bibnamefont {Parotto}}, \bibinfo {author}
  {\bibfnamefont {A.}~\bibnamefont {Pasztor}}, \bibinfo {author} {\bibfnamefont
  {C.}~\bibnamefont {Ratti}},\ and\ \bibinfo {author} {\bibfnamefont {K.~K.}\
  \bibnamefont {Szabo}},\ }\bibfield  {title} {\bibinfo {title} {{Resummed
  lattice QCD equation of state at finite baryon density: Strangeness
  neutrality and beyond}},\ }\href
  {https://doi.org/10.1103/PhysRevD.105.114504} {\bibfield  {journal} {\bibinfo
   {journal} {Phys. Rev. D}\ }\textbf {\bibinfo {volume} {105}},\ \bibinfo
  {pages} {114504} (\bibinfo {year} {2022}{\natexlab{b}})},\ \Eprint
  {https://arxiv.org/abs/2202.05574} {arXiv:2202.05574 [hep-lat]} \BibitemShut
  {NoStop}%
\bibitem [{\citenamefont {Fischer}(2019)}]{Fischer:2018sdj}%
  \BibitemOpen
  \bibfield  {author} {\bibinfo {author} {\bibfnamefont {C.~S.}\ \bibnamefont
  {Fischer}},\ }\bibfield  {title} {\bibinfo {title} {{QCD at finite
  temperature and chemical potential from Dyson--Schwinger equations}},\ }\href
  {https://doi.org/10.1016/j.ppnp.2019.01.002} {\bibfield  {journal} {\bibinfo
  {journal} {Prog. Part. Nucl. Phys.}\ }\textbf {\bibinfo {volume} {105}},\
  \bibinfo {pages} {1} (\bibinfo {year} {2019})},\ \Eprint
  {https://arxiv.org/abs/1810.12938} {arXiv:1810.12938 [hep-ph]} \BibitemShut
  {NoStop}%
\bibitem [{\citenamefont {Fischer}\ \emph {et~al.}(2014)\citenamefont
  {Fischer}, \citenamefont {Luecker},\ and\ \citenamefont
  {Welzbacher}}]{Fischer:2014ata}%
  \BibitemOpen
  \bibfield  {author} {\bibinfo {author} {\bibfnamefont {C.~S.}\ \bibnamefont
  {Fischer}}, \bibinfo {author} {\bibfnamefont {J.}~\bibnamefont {Luecker}},\
  and\ \bibinfo {author} {\bibfnamefont {C.~A.}\ \bibnamefont {Welzbacher}},\
  }\bibfield  {title} {\bibinfo {title} {{Phase structure of three and four
  flavor QCD}},\ }\href {https://doi.org/10.1103/PhysRevD.90.034022} {\bibfield
   {journal} {\bibinfo  {journal} {Phys. Rev. D}\ }\textbf {\bibinfo {volume}
  {90}},\ \bibinfo {pages} {034022} (\bibinfo {year} {2014})},\ \Eprint
  {https://arxiv.org/abs/1405.4762} {arXiv:1405.4762 [hep-ph]} \BibitemShut
  {NoStop}%
\bibitem [{\citenamefont {Isserstedt}\ \emph {et~al.}(2019)\citenamefont
  {Isserstedt}, \citenamefont {Buballa}, \citenamefont {Fischer},\ and\
  \citenamefont {Gunkel}}]{Isserstedt:2019pgx}%
  \BibitemOpen
  \bibfield  {author} {\bibinfo {author} {\bibfnamefont {P.}~\bibnamefont
  {Isserstedt}}, \bibinfo {author} {\bibfnamefont {M.}~\bibnamefont {Buballa}},
  \bibinfo {author} {\bibfnamefont {C.~S.}\ \bibnamefont {Fischer}},\ and\
  \bibinfo {author} {\bibfnamefont {P.~J.}\ \bibnamefont {Gunkel}},\ }\bibfield
   {title} {\bibinfo {title} {{Baryon number fluctuations in the QCD phase
  diagram from Dyson--Schwinger equations}},\ }\href
  {https://doi.org/10.1103/PhysRevD.100.074011} {\bibfield  {journal} {\bibinfo
   {journal} {Phys. Rev. D}\ }\textbf {\bibinfo {volume} {100}},\ \bibinfo
  {pages} {074011} (\bibinfo {year} {2019})},\ \Eprint
  {https://arxiv.org/abs/1906.11644} {arXiv:1906.11644 [hep-ph]} \BibitemShut
  {NoStop}%
\bibitem [{\citenamefont {Fu}\ \emph {et~al.}(2020)\citenamefont {Fu},
  \citenamefont {Pawlowski},\ and\ \citenamefont {Rennecke}}]{Fu:2019hdw}%
  \BibitemOpen
  \bibfield  {author} {\bibinfo {author} {\bibfnamefont {W.-j.}\ \bibnamefont
  {Fu}}, \bibinfo {author} {\bibfnamefont {J.~M.}\ \bibnamefont {Pawlowski}},\
  and\ \bibinfo {author} {\bibfnamefont {F.}~\bibnamefont {Rennecke}},\
  }\bibfield  {title} {\bibinfo {title} {{QCD phase structure at finite
  temperature and density}},\ }\href
  {https://doi.org/10.1103/PhysRevD.101.054032} {\bibfield  {journal} {\bibinfo
   {journal} {Phys. Rev. D}\ }\textbf {\bibinfo {volume} {101}},\ \bibinfo
  {pages} {054032} (\bibinfo {year} {2020})},\ \Eprint
  {https://arxiv.org/abs/1909.02991} {arXiv:1909.02991 [hep-ph]} \BibitemShut
  {NoStop}%
\bibitem [{\citenamefont {Gao}\ and\ \citenamefont
  {Pawlowski}(2020)}]{Gao:2020qsj}%
  \BibitemOpen
  \bibfield  {author} {\bibinfo {author} {\bibfnamefont {F.}~\bibnamefont
  {Gao}}\ and\ \bibinfo {author} {\bibfnamefont {J.~M.}\ \bibnamefont
  {Pawlowski}},\ }\bibfield  {title} {\bibinfo {title} {{QCD phase structure
  from functional methods}},\ }\href
  {https://doi.org/10.1103/PhysRevD.102.034027} {\bibfield  {journal} {\bibinfo
   {journal} {Phys. Rev. D}\ }\textbf {\bibinfo {volume} {102}},\ \bibinfo
  {pages} {034027} (\bibinfo {year} {2020})},\ \Eprint
  {https://arxiv.org/abs/2002.07500} {arXiv:2002.07500 [hep-ph]} \BibitemShut
  {NoStop}%
\bibitem [{\citenamefont {Gao}\ and\ \citenamefont
  {Pawlowski}(2021)}]{Gao:2020fbl}%
  \BibitemOpen
  \bibfield  {author} {\bibinfo {author} {\bibfnamefont {F.}~\bibnamefont
  {Gao}}\ and\ \bibinfo {author} {\bibfnamefont {J.~M.}\ \bibnamefont
  {Pawlowski}},\ }\bibfield  {title} {\bibinfo {title} {{Chiral phase structure
  and critical end point in QCD}},\ }\href
  {https://doi.org/10.1016/j.physletb.2021.136584} {\bibfield  {journal}
  {\bibinfo  {journal} {Phys. Lett. B}\ }\textbf {\bibinfo {volume} {820}},\
  \bibinfo {pages} {136584} (\bibinfo {year} {2021})},\ \Eprint
  {https://arxiv.org/abs/2010.13705} {arXiv:2010.13705 [hep-ph]} \BibitemShut
  {NoStop}%
\bibitem [{\citenamefont {Gunkel}\ and\ \citenamefont
  {Fischer}(2021)}]{Gunkel:2021oya}%
  \BibitemOpen
  \bibfield  {author} {\bibinfo {author} {\bibfnamefont {P.~J.}\ \bibnamefont
  {Gunkel}}\ and\ \bibinfo {author} {\bibfnamefont {C.~S.}\ \bibnamefont
  {Fischer}},\ }\bibfield  {title} {\bibinfo {title} {{Locating the critical
  endpoint of QCD: Mesonic backcoupling effects}},\ }\href
  {https://doi.org/10.1103/PhysRevD.104.054022} {\bibfield  {journal} {\bibinfo
   {journal} {Phys. Rev. D}\ }\textbf {\bibinfo {volume} {104}},\ \bibinfo
  {pages} {054022} (\bibinfo {year} {2021})},\ \Eprint
  {https://arxiv.org/abs/2106.08356} {arXiv:2106.08356 [hep-ph]} \BibitemShut
  {NoStop}%
\bibitem [{\citenamefont {Braun}\ \emph {et~al.}(2011)\citenamefont {Braun},
  \citenamefont {Haas}, \citenamefont {Marhauser},\ and\ \citenamefont
  {Pawlowski}}]{Braun:2009gm}%
  \BibitemOpen
  \bibfield  {author} {\bibinfo {author} {\bibfnamefont {J.}~\bibnamefont
  {Braun}}, \bibinfo {author} {\bibfnamefont {L.~M.}\ \bibnamefont {Haas}},
  \bibinfo {author} {\bibfnamefont {F.}~\bibnamefont {Marhauser}},\ and\
  \bibinfo {author} {\bibfnamefont {J.~M.}\ \bibnamefont {Pawlowski}},\
  }\bibfield  {title} {\bibinfo {title} {{Phase Structure of Two-Flavor QCD at
  Finite Chemical Potential}},\ }\href
  {https://doi.org/10.1103/PhysRevLett.106.022002} {\bibfield  {journal}
  {\bibinfo  {journal} {Phys. Rev. Lett.}\ }\textbf {\bibinfo {volume} {106}},\
  \bibinfo {pages} {022002} (\bibinfo {year} {2011})},\ \Eprint
  {https://arxiv.org/abs/0908.0008} {arXiv:0908.0008 [hep-ph]} \BibitemShut
  {NoStop}%
\bibitem [{\citenamefont {Fischer}\ \emph {et~al.}(2015)\citenamefont
  {Fischer}, \citenamefont {Luecker},\ and\ \citenamefont
  {Pawlowski}}]{Fischer:2014vxa}%
  \BibitemOpen
  \bibfield  {author} {\bibinfo {author} {\bibfnamefont {C.~S.}\ \bibnamefont
  {Fischer}}, \bibinfo {author} {\bibfnamefont {J.}~\bibnamefont {Luecker}},\
  and\ \bibinfo {author} {\bibfnamefont {J.~M.}\ \bibnamefont {Pawlowski}},\
  }\bibfield  {title} {\bibinfo {title} {{Phase structure of QCD for heavy
  quarks}},\ }\href {https://doi.org/10.1103/PhysRevD.91.014024} {\bibfield
  {journal} {\bibinfo  {journal} {Phys. Rev. D}\ }\textbf {\bibinfo {volume}
  {91}},\ \bibinfo {pages} {014024} (\bibinfo {year} {2015})},\ \Eprint
  {https://arxiv.org/abs/1409.8462} {arXiv:1409.8462 [hep-ph]} \BibitemShut
  {NoStop}%
\bibitem [{\citenamefont {Fischer}\ \emph {et~al.}(2010)\citenamefont
  {Fischer}, \citenamefont {Maas},\ and\ \citenamefont
  {Mueller}}]{Fischer:2010fx}%
  \BibitemOpen
  \bibfield  {author} {\bibinfo {author} {\bibfnamefont {C.~S.}\ \bibnamefont
  {Fischer}}, \bibinfo {author} {\bibfnamefont {A.}~\bibnamefont {Maas}},\ and\
  \bibinfo {author} {\bibfnamefont {J.~A.}\ \bibnamefont {Mueller}},\
  }\bibfield  {title} {\bibinfo {title} {{Chiral and deconfinement transition
  from correlation functions: SU(2) vs.\ SU(3)}},\ }\href
  {https://doi.org/10.1140/epjc/s10052-010-1343-1} {\bibfield  {journal}
  {\bibinfo  {journal} {Eur. Phys. J. C}\ }\textbf {\bibinfo {volume} {68}},\
  \bibinfo {pages} {165} (\bibinfo {year} {2010})},\ \Eprint
  {https://arxiv.org/abs/1003.1960} {arXiv:1003.1960 [hep-ph]} \BibitemShut
  {NoStop}%
\bibitem [{\citenamefont {Maas}\ \emph {et~al.}(2012)\citenamefont {Maas},
  \citenamefont {Pawlowski}, \citenamefont {{von Smekal}},\ and\ \citenamefont
  {Spielmann}}]{Maas:2011ez}%
  \BibitemOpen
  \bibfield  {author} {\bibinfo {author} {\bibfnamefont {A.}~\bibnamefont
  {Maas}}, \bibinfo {author} {\bibfnamefont {J.~M.}\ \bibnamefont {Pawlowski}},
  \bibinfo {author} {\bibfnamefont {L.}~\bibnamefont {{von Smekal}}},\ and\
  \bibinfo {author} {\bibfnamefont {D.}~\bibnamefont {Spielmann}},\ }\bibfield
  {title} {\bibinfo {title} {{The gluon propagator close to criticality}},\
  }\href {https://doi.org/10.1103/PhysRevD.85.034037} {\bibfield  {journal}
  {\bibinfo  {journal} {Phys. Rev. D}\ }\textbf {\bibinfo {volume} {85}},\
  \bibinfo {pages} {034037} (\bibinfo {year} {2012})},\ \Eprint
  {https://arxiv.org/abs/1110.6340} {arXiv:1110.6340 [hep-lat]} \BibitemShut
  {NoStop}%
\bibitem [{\citenamefont {Eichmann}\ \emph {et~al.}(2016)\citenamefont
  {Eichmann}, \citenamefont {Fischer},\ and\ \citenamefont
  {Welzbacher}}]{Eichmann:2015kfa}%
  \BibitemOpen
  \bibfield  {author} {\bibinfo {author} {\bibfnamefont {G.}~\bibnamefont
  {Eichmann}}, \bibinfo {author} {\bibfnamefont {C.~S.}\ \bibnamefont
  {Fischer}},\ and\ \bibinfo {author} {\bibfnamefont {C.~A.}\ \bibnamefont
  {Welzbacher}},\ }\bibfield  {title} {\bibinfo {title} {{Baryon effects on the
  location of QCD's critical end point}},\ }\href
  {https://doi.org/10.1103/PhysRevD.93.034013} {\bibfield  {journal} {\bibinfo
  {journal} {Phys. Rev. D}\ }\textbf {\bibinfo {volume} {93}},\ \bibinfo
  {pages} {034013} (\bibinfo {year} {2016})},\ \Eprint
  {https://arxiv.org/abs/1509.02082} {arXiv:1509.02082 [hep-ph]} \BibitemShut
  {NoStop}%
\bibitem [{\citenamefont {Ball}\ and\ \citenamefont
  {Chiu}(1980)}]{Ball:1980ay}%
  \BibitemOpen
  \bibfield  {author} {\bibinfo {author} {\bibfnamefont {J.~S.}\ \bibnamefont
  {Ball}}\ and\ \bibinfo {author} {\bibfnamefont {T.-W.}\ \bibnamefont
  {Chiu}},\ }\bibfield  {title} {\bibinfo {title} {{Analytic properties of the
  vertex function in gauge theories. I}},\ }\href
  {https://doi.org/10.1103/PhysRevD.22.2542} {\bibfield  {journal} {\bibinfo
  {journal} {Phys. Rev. D}\ }\textbf {\bibinfo {volume} {22}},\ \bibinfo
  {pages} {2542} (\bibinfo {year} {1980})}\BibitemShut {NoStop}%
\bibitem [{\citenamefont {Gunkel}\ \emph {et~al.}(2019)\citenamefont {Gunkel},
  \citenamefont {Fischer},\ and\ \citenamefont {Isserstedt}}]{Gunkel:2019xnh}%
  \BibitemOpen
  \bibfield  {author} {\bibinfo {author} {\bibfnamefont {P.~J.}\ \bibnamefont
  {Gunkel}}, \bibinfo {author} {\bibfnamefont {C.~S.}\ \bibnamefont
  {Fischer}},\ and\ \bibinfo {author} {\bibfnamefont {P.}~\bibnamefont
  {Isserstedt}},\ }\bibfield  {title} {\bibinfo {title} {{Quarks and light
  (pseudo-)scalar mesons at finite chemical potential}},\ }\href
  {https://doi.org/10.1140/epja/i2019-12868-1} {\bibfield  {journal} {\bibinfo
  {journal} {Eur. Phys. J. A}\ }\textbf {\bibinfo {volume} {55}},\ \bibinfo
  {pages} {169} (\bibinfo {year} {2019})},\ \Eprint
  {https://arxiv.org/abs/1907.08110} {arXiv:1907.08110 [hep-ph]} \BibitemShut
  {NoStop}%
\bibitem [{\citenamefont {Isserstedt}(2021)}]{Isserstedt:2021acw}%
  \BibitemOpen
  \bibfield  {author} {\bibinfo {author} {\bibfnamefont {P.}~\bibnamefont
  {Isserstedt}},\ }\emph {\bibinfo {title} {{Thermodynamics of
  strong-interaction matter: On the phase structure and thermodynamics of
  quantum chromodynamics with Dyson--Schwinger equations}}},\ \href
  {https://doi.org/10.22029/jlupub-310} {Ph.D. thesis},\ \bibinfo  {school}
  {Giessen University}, \bibinfo {address} {Germany} (\bibinfo {year}
  {2021})\BibitemShut {NoStop}%
\bibitem [{\citenamefont {Bonati}\ \emph {et~al.}(2015)\citenamefont {Bonati},
  \citenamefont {D'Elia}, \citenamefont {Mariti}, \citenamefont {Mesiti},
  \citenamefont {Negro},\ and\ \citenamefont {Sanfilippo}}]{Bonati:2015bha}%
  \BibitemOpen
  \bibfield  {author} {\bibinfo {author} {\bibfnamefont {C.}~\bibnamefont
  {Bonati}}, \bibinfo {author} {\bibfnamefont {M.}~\bibnamefont {D'Elia}},
  \bibinfo {author} {\bibfnamefont {M.}~\bibnamefont {Mariti}}, \bibinfo
  {author} {\bibfnamefont {M.}~\bibnamefont {Mesiti}}, \bibinfo {author}
  {\bibfnamefont {F.}~\bibnamefont {Negro}},\ and\ \bibinfo {author}
  {\bibfnamefont {F.}~\bibnamefont {Sanfilippo}},\ }\bibfield  {title}
  {\bibinfo {title} {Curvature of the chiral pseudocritical line in qcd:
  Continuum extrapolated results},\ }\href
  {https://doi.org/10.1103/PhysRevD.92.054503} {\bibfield  {journal} {\bibinfo
  {journal} {Phys. Rev. D}\ }\textbf {\bibinfo {volume} {92}},\ \bibinfo
  {pages} {054503} (\bibinfo {year} {2015})},\ \Eprint
  {https://arxiv.org/abs/1507.03571} {arXiv:1507.03571 [hep-lat]} \BibitemShut
  {NoStop}%
\bibitem [{\citenamefont {Bellwied}\ \emph {et~al.}(2015)\citenamefont
  {Bellwied} \emph {et~al.}}]{Bellwied:2015rza}%
  \BibitemOpen
  \bibfield  {author} {\bibinfo {author} {\bibfnamefont {R.}~\bibnamefont
  {Bellwied}} \emph {et~al.},\ }\bibfield  {title} {\bibinfo {title} {{The QCD
  phase diagram from analytic continuation}},\ }\href
  {https://doi.org/10.1016/j.physletb.2015.11.011} {\bibfield  {journal}
  {\bibinfo  {journal} {Phys. Lett. B}\ }\textbf {\bibinfo {volume} {751}},\
  \bibinfo {pages} {559} (\bibinfo {year} {2015})},\ \Eprint
  {https://arxiv.org/abs/1507.07510} {arXiv:1507.07510 [hep-lat]} \BibitemShut
  {NoStop}%
\bibitem [{\citenamefont {Bonati}\ \emph {et~al.}(2018)\citenamefont {Bonati},
  \citenamefont {D'Elia}, \citenamefont {Negro}, \citenamefont {Sanfilippo},\
  and\ \citenamefont {Zambello}}]{Bonati:2018nut}%
  \BibitemOpen
  \bibfield  {author} {\bibinfo {author} {\bibfnamefont {C.}~\bibnamefont
  {Bonati}}, \bibinfo {author} {\bibfnamefont {M.}~\bibnamefont {D'Elia}},
  \bibinfo {author} {\bibfnamefont {F.}~\bibnamefont {Negro}}, \bibinfo
  {author} {\bibfnamefont {F.}~\bibnamefont {Sanfilippo}},\ and\ \bibinfo
  {author} {\bibfnamefont {K.}~\bibnamefont {Zambello}},\ }\bibfield  {title}
  {\bibinfo {title} {Curvature of the pseudocritical line in qcd: Taylor
  expansion matches analytic continuation},\ }\href
  {https://doi.org/10.1103/PhysRevD.98.054510} {\bibfield  {journal} {\bibinfo
  {journal} {Phys. Rev. D}\ }\textbf {\bibinfo {volume} {98}},\ \bibinfo
  {pages} {054510} (\bibinfo {year} {2018})},\ \Eprint
  {https://arxiv.org/abs/1805.02960} {arXiv:1805.02960 [hep-lat]} \BibitemShut
  {NoStop}%
\bibitem [{\citenamefont {Bazavov}\ \emph
  {et~al.}(2019{\natexlab{b}})\citenamefont {Bazavov} \emph
  {et~al.}}]{HotQCD:2018pds}%
  \BibitemOpen
  \bibfield  {author} {\bibinfo {author} {\bibfnamefont {A.}~\bibnamefont
  {Bazavov}} \emph {et~al.} (\bibinfo {collaboration} {HotQCD}),\ }\bibfield
  {title} {\bibinfo {title} {Chiral crossover in qcd at zero and non-zero
  chemical potentials},\ }\href
  {https://doi.org/10.1016/j.physletb.2019.05.013} {\bibfield  {journal}
  {\bibinfo  {journal} {Phys. Lett. B}\ }\textbf {\bibinfo {volume} {795}},\
  \bibinfo {pages} {15} (\bibinfo {year} {2019}{\natexlab{b}})},\ \Eprint
  {https://arxiv.org/abs/1812.08235} {arXiv:1812.08235 [hep-lat]} \BibitemShut
  {NoStop}%
\bibitem [{\citenamefont {Binosi}\ \emph {et~al.}(2009)\citenamefont {Binosi},
  \citenamefont {Collins}, \citenamefont {Kaufhold},\ and\ \citenamefont
  {Theussl}}]{Binosi:2008ig}%
  \BibitemOpen
  \bibfield  {author} {\bibinfo {author} {\bibfnamefont {D.}~\bibnamefont
  {Binosi}}, \bibinfo {author} {\bibfnamefont {J.}~\bibnamefont {Collins}},
  \bibinfo {author} {\bibfnamefont {C.}~\bibnamefont {Kaufhold}},\ and\
  \bibinfo {author} {\bibfnamefont {L.}~\bibnamefont {Theussl}},\ }\bibfield
  {title} {\bibinfo {title} {{Jaxo\-Draw: A graphical user interface for
  drawing Feynman diagrams. Version 2.0 release notes}},\ }\href
  {https://doi.org/10.1016/j.cpc.2009.02.020} {\bibfield  {journal} {\bibinfo
  {journal} {Comput. Phys. Commun.}\ }\textbf {\bibinfo {volume} {180}},\
  \bibinfo {pages} {1709} (\bibinfo {year} {2009})},\ \Eprint
  {https://arxiv.org/abs/0811.4113} {arXiv:0811.4113 [hep-ph]} \BibitemShut
  {NoStop}%
\end{thebibliography}%

\end{document}